\documentclass[11pt,twocolumn,twoside]{osajnl}

\journal{ol} 

\setboolean{shortarticle}{true}

\title{Vectorial dark dissipative solitons in Kerr resonators}

\author[1]{B. Kostet}
\author[1,4]{S. S. Gopalakrishnan}
\author[1]{E. Averlant}
\author[1]{Y. Soupart}
\author[2,3]{K. Panajotov}
\author[1,*]{M. Tlidi}
\affil[*]{Corresponding author: mtlidi@ulb.ac.be}

\affil[1]{Facult\'{e} des Sciences, Universit\'{e} libre de Bruxelles (ULB), CP. 231, 1050 Brussels, Belgium}
\affil[2]{Department of Applied Physics and Photonics (IR-TONA), Vrije Universiteit Brussels, Pleinlaan 2, 1050 Brussels, Belgium}
\affil[3]{Institute of Solid State Physics, 72 Tzarigradsko CHaussee Blvd., 1784 Sofia, Bulgaria}
\affil[4]{Laboratoire de Physique des Lasers, Atomes et Mol\'{e}cules, CNRS UMR 8523, Universit\'{e} Lille 1 - 59655 Villeneuve d'Ascq Cedex, France}

\begin{abstract}
	We report the existence of vectorial dark dissipative solitons in optical cavities subject to a coherently injected beam. 
	We assume that the resonator is operating in a normal dispersion regime far from any modulational instability. 
	We show that the vectorial front locking mechanism allows for the stabilisation of dark dissipative structures. These structures differ by their temporal duration and their state of polarization. We characterize them by constructing their heteroclinic snaking bifurcation diagram showing evidence of multistability within a finite range of the control parameter.
\end{abstract}

\setboolean{displaycopyright}{true}

\begin{document}
	
	\maketitle
	
	The formation of dissipative solitons (DSs) and their link to optical frequency comb generation in coherently-driven dispersive Kerr resonators has been established  in  \cite{matsko2011mode,coen2013moctave,refId0,lugiato2018kenberberg}. Optical frequency comb generation in these devices can be seen as the spectral content of DSs. Besides their impact on fundamental physics, optical frequency combs have direct applications in precision distance measurements, optical waveform, and
	optical spectroscopy \cite{ferdous2011spectral,kippenberg2011microresonator}. The formation of scalar, i.e., in the absence of polarization degrees of freedom, DSs have been abundantly discussed and is by now fairly well understood  \cite{matsko2011mode,coen2013moctave,refId0,lugiato2018kenberberg}.

	When considering polarization degree of freedom due to the birefringence of the Kerr media, vectorial DS can have a nontrivial 
	polarization state, resulting in a richer dynamics.  Modulational Instability (MI)   \cite{haelterman1994polarization} can lead to a spontaneous symmetry breaking of DS  \cite{xu2020spontaneous}, group velocity-locked vector solitons \cite{sergeyev2014spiral,mou2011all} or polarization locked vector soliton \cite{akhmediev1994elliptically,cundiff1999observation,Sanchez-Morcillo_00,garbin2020dissipative}. More recently, the existence of dark-bright vector solitons arising from a cross-polarization coupling in birefringent fiber lasers have been experimentally shown in \cite{hu2020dissipative}.  The coexistence of two different types of bright vectorial DS in driven  Kerr resonators was theoretically predicted \cite{averlant2017coexistence,suzuki2018theoretical,Saha2020}, and has been recently experimentally observed \cite{copie2019interplay,nielsen2019coexistence}.  All the above mentioned studies have been obtained in single mode operation assuming the mean field limit. In contrast, other coexisting DSs without polarization degrees of freedom have been reported theoretically \cite{hansson2015frequency}, and experimentally, in regimes where the Kerr resonator exhibits multimode operation \cite{anderson2017coexistence}.

	The aim of this letter is to investigate the polarization properties of dark vectorial dissipative solitons (DVDSs) in Kerr resonators operating in a normal dispersion regime far from any MI. The mechanism of their formation is attributed to the locking of fronts connecting two different continuous-wave (CW) states.  They are characterized by different widths, different polarization properties, and can coexist in a finite range of the system parameters.  We characterize these findings through a bifurcation analysis, and show that DVDSs exhibit a heteroclinic snaking type of behavior. This behavior has been established in the scalar case \cite{parra2016dark}, and it is a well documented issue in other  systems operating far from equilibrium \cite{CoulletPRL1989,KNOBLOCH200582}. The effect of polarization-induced cross-phase modulation on MI patterns has been studied in \cite{Hansson:18}. Note that front dynamics or switching waves \cite{coen1999convection}, and dark dissipative solitons \cite{garbin2017experimental} have been experimentally investigated in the scalar case, i.e., without the polarization degree of freedom due to the inherent birefringence of the $\chi^{(3)}$ material. 
	
	Vectorial driven Kerr fiber resonators, or microcavities are described by two-coupled  Lugiato--Lefever equations (LLEs) \cite{haelterman1994polarization,Hansson:18} 
		\begin{eqnarray}
			\label{eq:LLV}
			\partial_{t}E_{x,y} &=& E_{Ix,Iy}+ i\left(|E_{x,y}|^2 + b |E_{y,x}|^2\right)E_{x,y}\\ \nonumber
			&-&\left(1+i\theta_{x,y}\mp \Delta\beta_{1}\partial_{\tau}-i\beta_2 \partial_{\tau\tau}\right) E_{x,y}.
		\end{eqnarray}
	Here $E_{x,y}$ are the slowly varying envelopes of the field components polarized respectively along the slow and the fast axis of the birefringent fiber. $E_{Ix,Iy}=E_I(\cos{\psi}, \sin{\psi})$ is  the injected beam amplitude which has a linear polarization that forms an angle $\psi$ with respect to the slow axis. We fix the value of $\psi$ to $\pi/4$, $\theta_{x,y}$ are the frequency detunings between the driven field and the cavity resonances in the corresponding polarization directions. We consider a weakly birefringent optical fiber cavity where the cross-phase modulation coefficient is $b=2/3$. The last term in Eq.~\ref{eq:LLV} describes the normal chromatic dispersion with $\tau$ being the fast time variable in the reference frame moving with the mean group velocity of the two polarization fields. The slow time variable $t$ describes the evolution of the envelope over cavity round-trips. 
	The difference between the first order coefficients $\Delta \beta_1 = (\beta_{1x}-\beta_{1y})/2$ determines the group velocity mismatch between the two polarization components. We will assume it to be zero in our study, as such an approximation holds in practical systems \cite{nielsen2019coexistence,6720193}.
	
	We focus on the bistable regime where the homogeneous steady or continuous-wave (CW) states are modulationally stable. Fig.~\ref{fig:HSS}a shows the stability map in the ($\theta_x$, $E_I$) parameter space. We fix $\theta_x$ and $\theta_y$, and let the injection strength be the control parameter. The output input characteristics corresponding to CW solutions is shown in Fig.\ref{fig:HSS}b. It evidences that the cavity operates far from any MI. In this regime, vectorial fronts connecting the upper and the lower CW states can be generated which are shown by the total intensity $S_0=|E_x|^2+|E_y|^2$ temporal profiles in Fig.\ref{fig:HSS}c. For small injected power, the lower stable state invades the system resulting in a front propagating to the right (see Fig.~\ref{fig:HSS}c (left panel). By increasing the injected field intensity, we reach the so-called Maxwell point where the vectorial front is motionless as shown in Fig.~\ref{fig:HSS}c (middle panel). When further increasing the injected field intensity, the upper states invades the system and the vectorial front propagates to its left as shown in Fig.\ref{fig:HSS}(c) (right panel).
Unlike in systems where domain walls are formed leading to the formation of DVDSs \cite{garbin2020dissipative} obtained for $b>1$, here for $b=2/3$,  the total intensity consists of a vectorial front that is devoid of any topological defects. 
\begin{figure}
	\unitlength=35.0mm
	\centerline{
		\includegraphics[width=\linewidth]{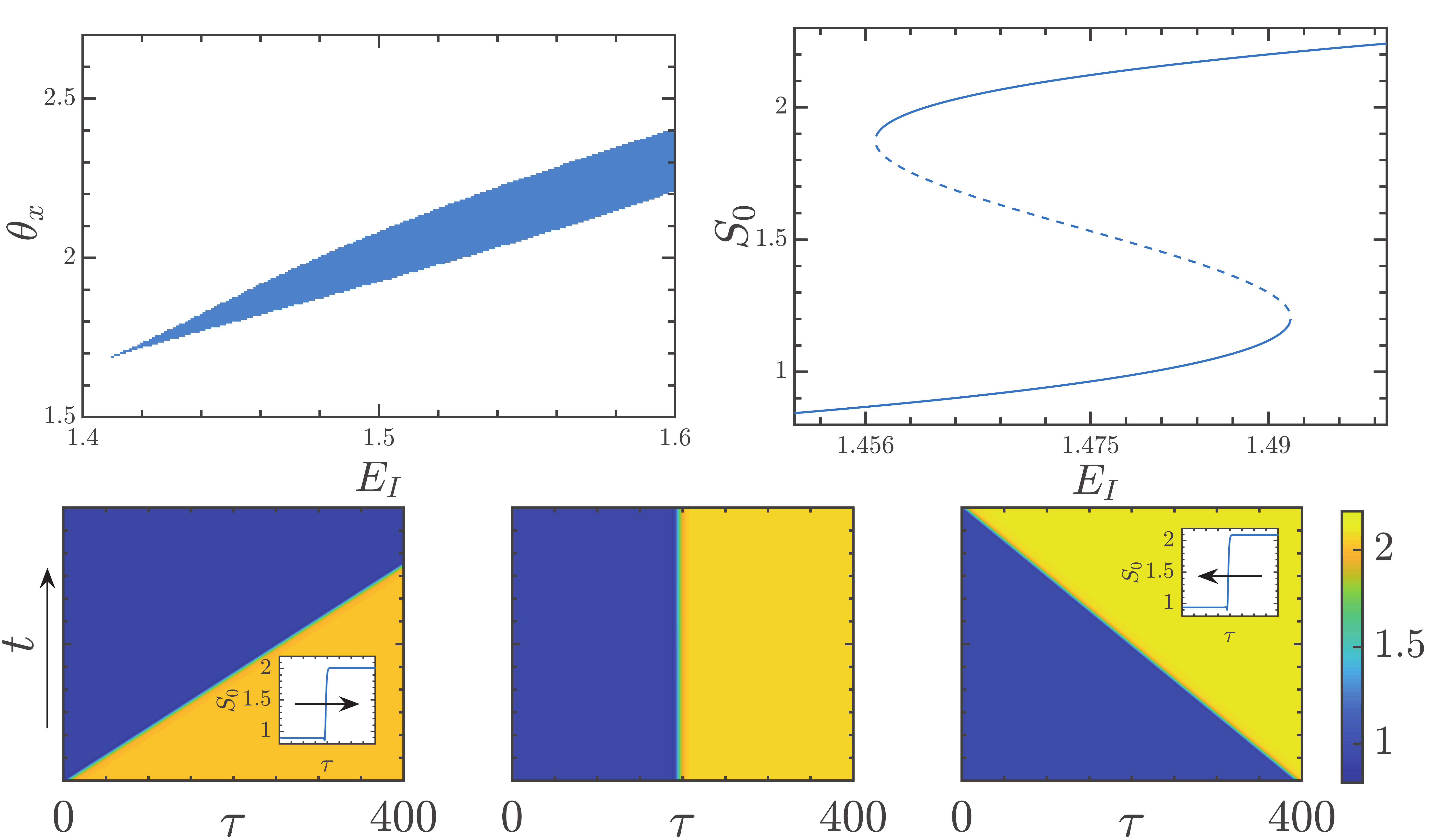}
	}
	\begin{picture}(0,0)
		\put(-0.1,1.4){(a)}
		\put(1.2,1.4){(b)}
		\put(-0.1,0.65){(c)}
	\end{picture}
	\caption{(a) Stability map in the $\theta_x$-$E_I$ plane, with $\theta_y= 1.95$. The white (blue) region indicates monostability (bistability).\\
		(b) Bistable curve obtained for $\theta_x=1.90$ and  $\theta_y=1.95$. Stable (unstable) CW states are denoted by solid (dashed) lines.\\ 
		(c) Front propagation to the right, below the Maxwell point (left panel: $E_I = 1.462$), at the Maxwell point (middle panel: $E_I = 1.46652$), and above the Maxwell point (right panel: $E_I = 1.471$).
		The boundary conditions are fixed to the CW values.}
	\label{fig:HSS}
\end{figure}
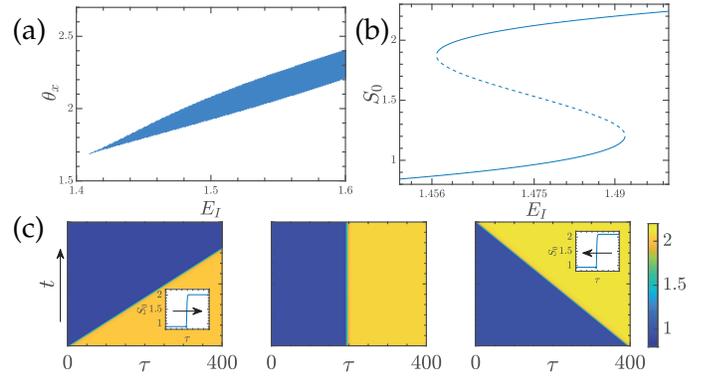 
\begin{figure}
	\unitlength=65.0mm
	\centerline{
		\includegraphics[width=\linewidth]{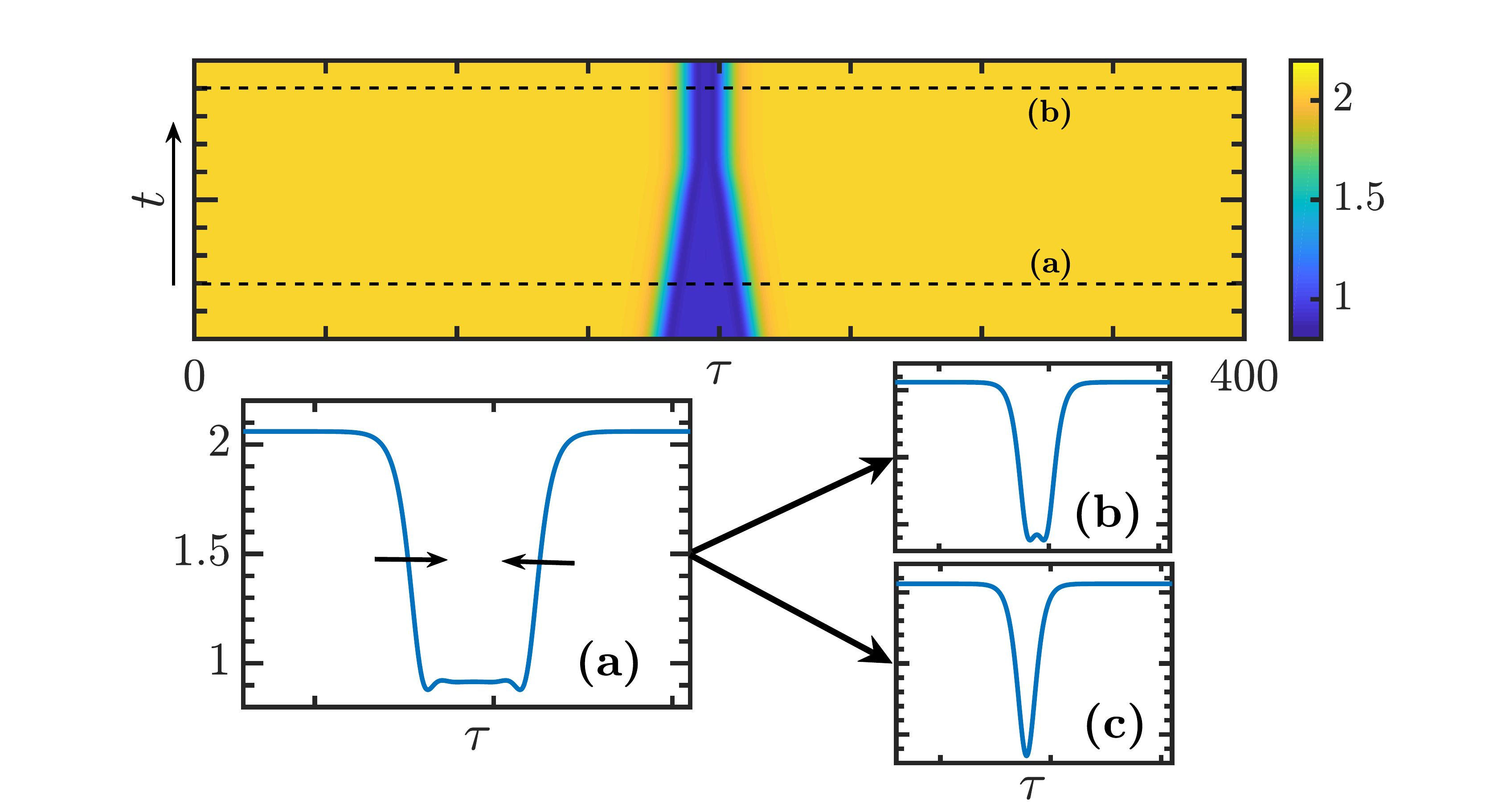}
	}
	\caption{Two fronts interact in an attractive way, which leads to the formation of vectorial dark dissipative solitons with the total intensity in the $\tau$-$t$ map (top panel).   (a,b) Cross-sections along the dashed lines indicated in the  $\tau$-$t$ map. (c) stable single dip DVDS. Numerical simulations of Eq.~\ref{eq:LLV} are obtained for the parameters $E_I = 1.4675$, $\theta_x = 1.90$, and  $\theta_y = 1.95$.
	}
	\label{fig:frontlocking}
\end{figure} 

\begin{figure}[ht]
	\unitlength=75.0mm
	\centerline{
		\includegraphics[width=\linewidth]{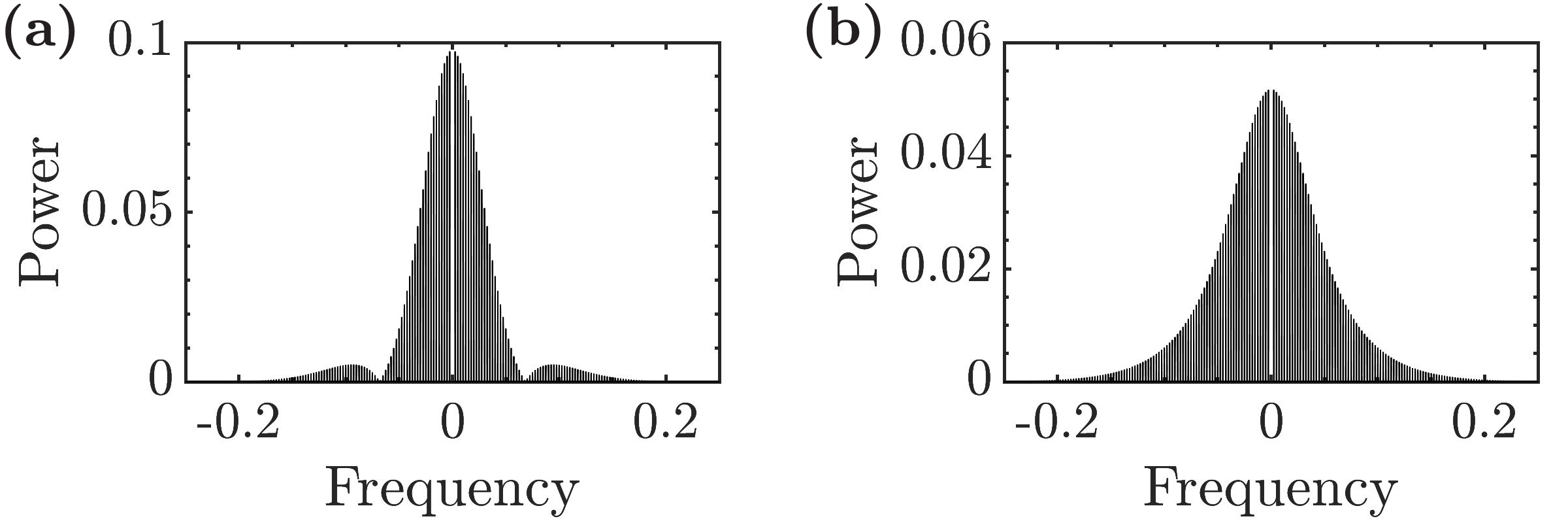}
	}
	\caption{Optical frequency combs corresponding to the two coexisting profiles shown in \ref{fig:frontlocking}. Parameters are $E_I = 1.46655$, $\theta_x = 1.90$, and $\theta_y = 1.95$.}
	
	\label{fig:combs}
\end{figure}

The profiles of the vectorial fronts shown in the insets of Fig.~\ref{fig:HSS}(c) and Fig.~\ref{fig:frontlocking}(a) evidence that they possess damped oscillatory tails on the lower CW state. In this case, the front interaction alternates between attractive and repulsive with an intensity that depends on the distance between fronts. As a consequence, the stabilization is attributed to front locking. If however, the front tails are exponentially decaying without damped oscillations, front interaction is always attractive and decays exponentially with the distance between the fronts. In this case, stable DS solutions are impossible \cite{CoulletPRL1989,KNOBLOCH200582,MA20101867}. The nature of the front interaction is dictated by their tails. Indeed, when two-well separated vectorial fronts propagate in the cavity, they interact and attract each other as shown by the temporal evolution of the total intensity  $S_0$  in the $(\tau,t)$ map of Fig.~\ref{fig:frontlocking}(a). Cross-sections of these vectorial fronts along the dashed lines denoted in Fig.~\ref{fig:frontlocking}(a) are shown in Fig.\ref{fig:frontlocking}(b) and Fig.~\ref{fig:frontlocking}(c).  As the fronts approach one another they start to interact via their damped oscillatory tails, attracting each other, and subsequently converging towards a stable DVDS. As can be seen in the map in the upper part of Fig.~\ref{fig:frontlocking}, after a long time evolution, the two fronts will interact and attract each other in the course of time until they reach a stationary stable DVDS state. Depending on the initial condition, the system can reach two different states consisting of either a dip with temporal oscillations (a small  bump) at its center [see Fig.~\ref{fig:frontlocking}(b)] or a single dip [see Fig.~\ref{fig:frontlocking}(c)]. In the case shown in this map, the fronts converge to form Fig.~\ref{fig:frontlocking}(b). These heteroclinic solutions resulting from front locking connection between CW solutions of the bistable response have different widths, and coexist for the same values of the system parameters. The spectral properties of these two structures are shown in Fig.~\ref{fig:combs}, indicating that they form two distincts combs with the same frequency line spacing. The combs Fig.~\ref{fig:combs}(a) and Fig.~\ref{fig:combs}(b) correspond respectively to the profiles Fig.~\ref{fig:frontlocking}(b) and Fig.~\ref{fig:frontlocking}(c).

\begin{figure}[ht]
	\unitlength=75.0mm
	\centerline{
		\includegraphics[width=\linewidth]{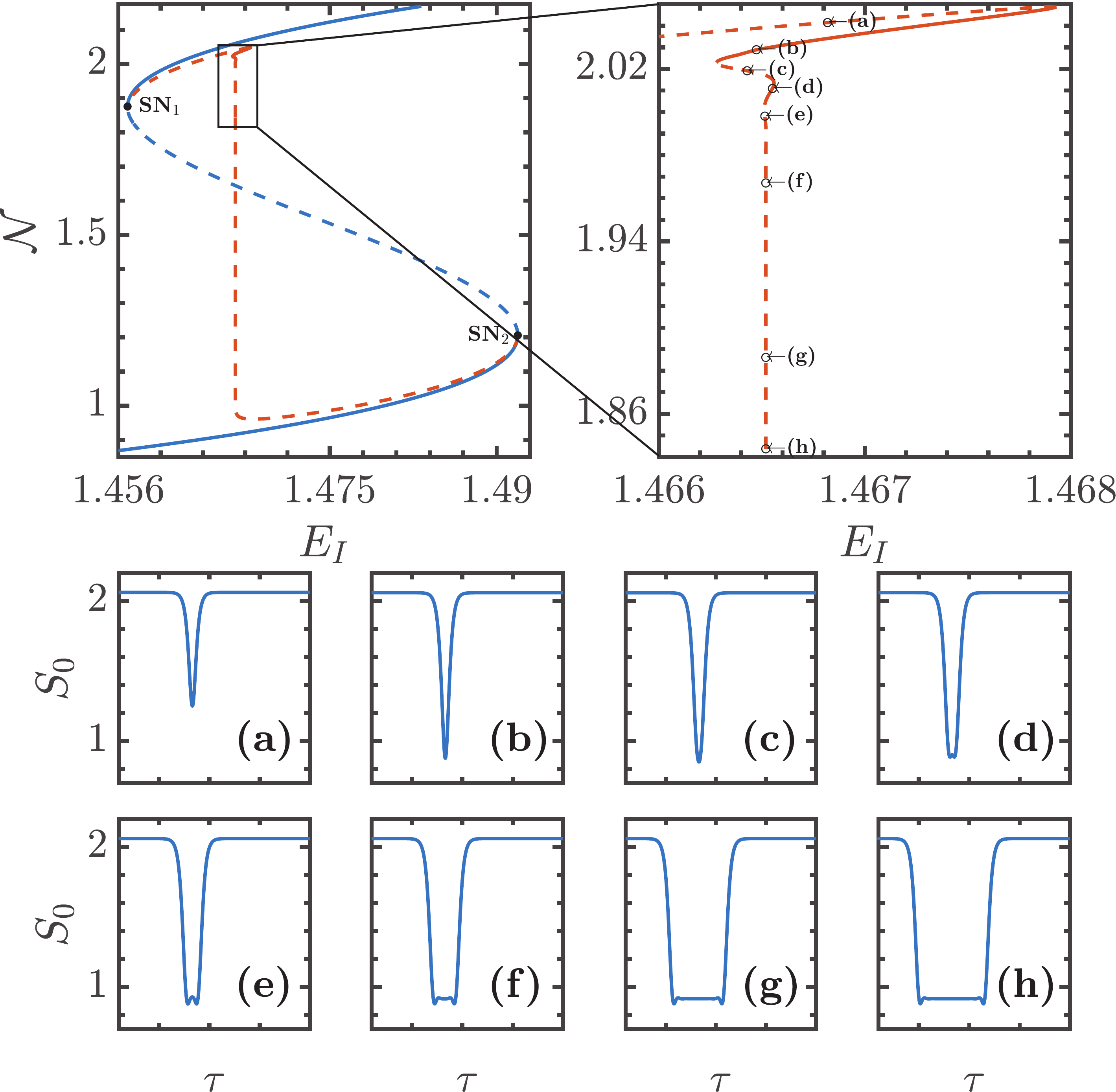}
	}
	\caption{Bifurcation diagram of Eq. (1) showing snaking curve corresponding vectorial dark dissipative solitons as a function of the injected field amplitude.   The right panel is a zoom around the snaking curve. Solid (dotted) curves correspond to stable (unstable) localized
		solutions.  (a--h) Profiles of the total field intensity $S_0$. Parameters are $\theta_x = 1.90$, $\theta_y = 1.95$.}
	\label{fig:snaking}
\end{figure} 
Since the intensity of DVDSs having different width are more or less similar, it
is convenient to plot the  “$L_2$ norm”,  defined as ${\cal{N}}=(\int_0^{L} S_0 \;\mathrm{d}\tau)/L$ as a function of the injected field amplitude. This yields the snaking curve shown in the bifurcation diagram of Fig.\ref{fig:snaking}.  As  ${\cal{N}}$ decreases, at each turning point where the slope becomes infinite, a pair of additional small bumps appear in the center of the DVDS. One sees that this behavior, referred to as
the collapsed heteroclinic snaking phenomenon \cite{CoulletPRL1989,KNOBLOCH200582,MA20101867}. The snaking curve is shown in the magnified close-up view shown alongside.  The snaking curve (orange curve) emerges from the saddle-node (denoted by SN$_1$ in Fig.~\ref{fig:snaking}) which connects the CW solutions (blue curve). We start from a known solution and move along the stable branch of DVDS (b) by slowly decreasing $E_I$. The total intensity profile corresponding to the stable DVDS is shown in panel (b).
When further decreasing the value of $E_I$, ${\cal{N}}$ decreases, and reaches a saddle-node bifurcation at the end of branch (b). By slowly increasing $E_I$ the solution is numerically continued along the unstable branch (c), which connects to another stable branch (d) via another saddle-node bifurcation. The width of the DVDS increases, and as the unstable branch connects to the stable branch (d), locked vectorial fronts develop a dip which stabilises the solitons. This bump is born when a stable branch connects to an unstable branch at a saddle-node bifurcation. Following this unstable branch (e) leads us to an oscillating curve whose amplitude dampens exponentially until it connects the saddle-node denoted by SN$_2$ in Fig.~\ref{fig:snaking}  associated with CW solutions. The different profiles of the DVDS along the different branches shown in the bifurcation diagram are presented in Fig.~\ref{fig:snaking} in panels (a--h). The different solutions indicate that the DVDS widens along the collapsing curve. 
\begin{figure}[ht]
	\unitlength=52.0mm
	\centerline{
		\includegraphics[width=\linewidth]{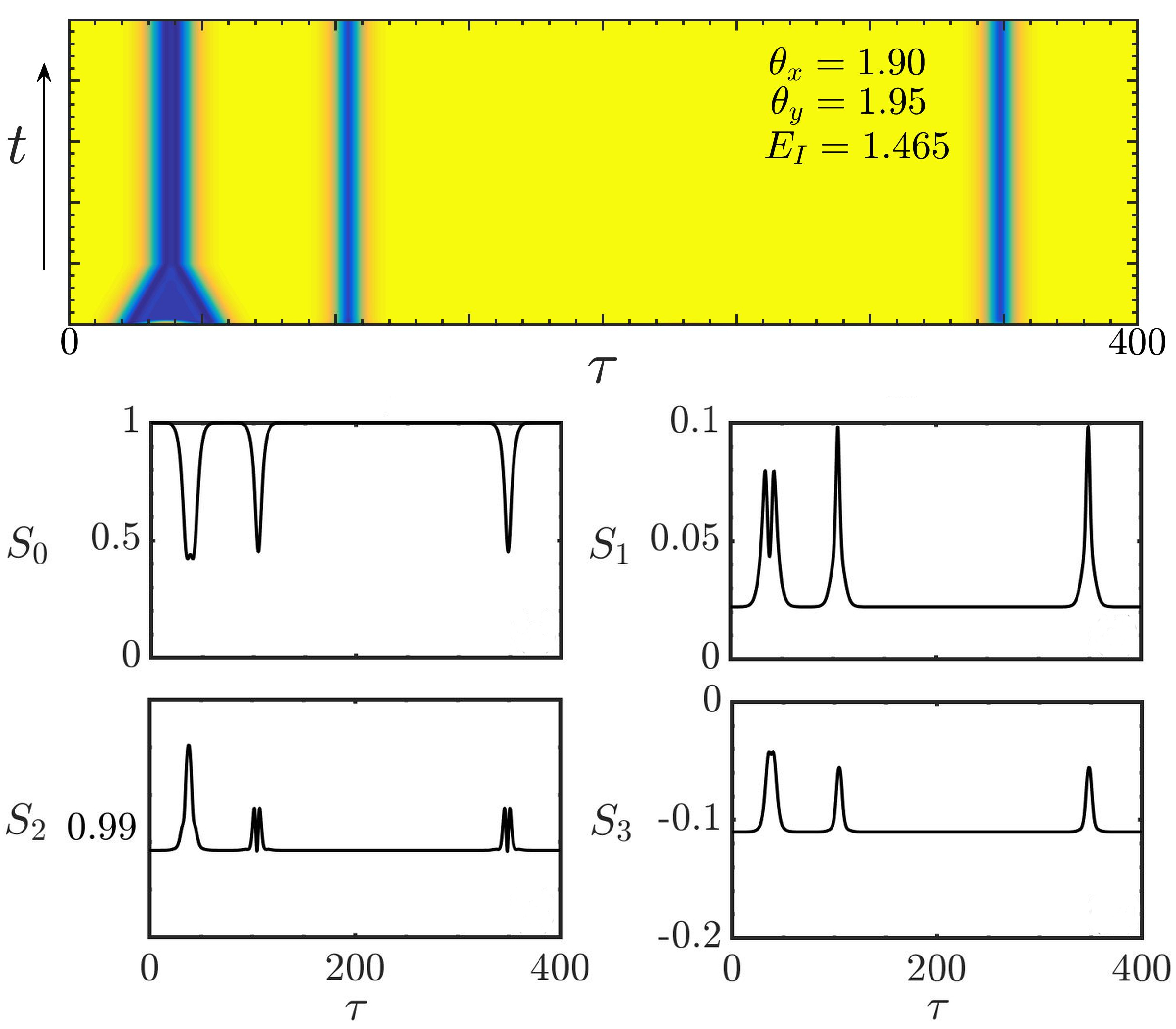}
	}
	\caption{Spontaneous generation of dark vectorial temporal dissipative solitons.\\ (Top) $(\tau,t)$ map of the total intensity $S_0$. Parameters are the same as in Fig.~\ref{fig:combs}.\\
		(Bottom) Normalized Stokes parameters $S_0, S_1, S_2$ and $S_3$ as a function of the fast time where ($^*$) stands for the complex conjugate.
	}
	\label{fig:SINGLEDARK}
\end{figure}

Finally, to show that the DVDS generation is robust, leading to stable structures, we perform a numerical simulation of Eq.~\ref{eq:LLV}. The initial condition was taken as a small random perturbation to a field with an amplitude of $1.5$ \big(a value in the middle of the two stable states shown in Fig.~\ref{fig:HSS}(a)\big), on a periodic domain. The time evolution is shown in the  $\tau$-$t$ map of Fig.~\ref{fig:SINGLEDARK}(a). The intracavity total intensity evolves towards stationary solutions consisting of two-different DVDSs with different temporal durations as shown in Fig.~\ref{fig:SINGLEDARK}. 

We emphasize the vectorial character of the dark dissipative solitons discussed here by presenting the temporal profiles of the normalized Stokes parameters, defined as $S_0=|E_x|^2+|E_y|^2$,  $S_1=(|E_x|^2-|E_y|^2)/S_0$,  $S_2=2\textrm{Re}(E_xE_y^*)/S_0$, $S_3=-2i(E_xE_y^*)/S_0$, in the bottom panel of Fig.~\ref{fig:SINGLEDARK}. There is a clear deviation of the DVDS polarization state from the one of the injected light $E_I$ (linearly polarized at 45 degrees with $S_{I1}=S_{I3}=0$ and $S_{I2}=1$). The polarization of the dark vector soliton changes along its profile and acquires well pronounced ellipticity with $S_3\neq 0$. Additionally, we can notice that there is a small difference between the polarization properties of the two different solutions.

In summary, we have demonstrated that robust vectorial dark dissipative solitons can be hosted in Kerr cavities without any modulational instability but in normal dispersion regime. We identify a family of dark vectorial dissipative solitons. They differ by their temporal duration and their state of  polarization. We have characterized these solitons by establishing their collapsed snaking bifurcation diagram. Finally, we have shown that these structures emerge spontaneously from random fluctuations, and two of them can coexist in the same system.

Funding:  K.P. acknowledges the support by the Fonds Wetenschappelijk Onderzoek-Vlaanderen FWO (G0E5819N). M.T. is a Research Director with the Fonds de la Recherche Scientifique F.R.S.-FNRS, Belgium.

Disclosures. The authors declare no conflicts of interest.	
\bibliography{Dark_OL}
\bibliographyfullrefs{Dark_OL}	

\end{document}